\documentclass{optica-article}

\journal{opticajournal} 

\renewcommand{\mu}{\upmu}

\articletype{Research Article}

\usepackage{lineno}

\usepackage{comment}
\usepackage{float}  

\begin{document}

\title{Mid-infrared quantum scanning microscopy via visible light beyond spatial correlations}

\author{\author{Josué R. León-Torres,\authormark{1, 2, 3*} Vasile-Laurențiu Dosan,\authormark{1,4,5} Marija M. Ćurčić,\authormark{6} Alek Lagarrigue,\authormark{4}  Frank Setzpfandt,\authormark{1, 2}  Markus Gräfe,\authormark {2, 7} and Valerio Flavio Gili\authormark{2}}}


\address{
\authormark{1}Abbe Center of Photonics, Friedrich Schiller University Jena, Albert-Einstein-Straße 6, 07745 Jena, Germany\\

\authormark{2}Fraunhofer Institute for Applied Optics and Precision Engineering IOF, Albert-Einstein-Straße 7, 07745 Jena, Germany\\

\authormark{3}Cluster of Excellence Balance of the Microverse, Friedrich Schiller University Jena, Jena, Germany\\

\authormark{4}Quantum Optics Jena GmbH,  Am Zementwerk 8, 07745 Jena, Germany\\

\authormark{5}Max Planck School of Photonics, Friedrich Schiller University Jena, Albert-Einstein-Straße 15, 07745 Jena, Germany

\authormark{6}Institute of Physics Belgrade, University of Belgrade, Pregrevica 118, 11080 Zemun, Belgrade, Serbia\\

\authormark{7}Institute for Applied Physics, Technical University of Darmstadt, Otto-Berndt-Straße 3, 64287 Darmstadt, Germany}

\email{\authormark{*}josue.ricardo.leon.torres@iof.fraunhofer.de} 

\begin{abstract*} 
The mid-infrared (MIR) region of the electromagnetic spectrum spans from 2- to 25-$\mu \mathrm{m}$, serving as a valuable tool for accessing rich chemical information. Functional groups, lipids, and other complex molecules can be analyzed by optical absorption measurements due to their vibrational modes in the MIR spectral region. 
Over the past few decades, this field has faced challenges due to difficulties in generating MIR light and the limited maturity of detection systems in this spectral range. Quantum imaging with undetected light (QIUL) provides a spectrally tuneable photon-pair source, in which the sample can be illuminated with MIR light while visible (VIS) light is employed for detection and image reconstruction, overcoming the detection limitations and benefiting from the rich chemical information of the MIR spectral region. All previous QIUL implementations are based on spatial correlations, 
which are never perfect and thus hindered the imaging performance. In this work, we implement a raster-scanning QIUL method that is independent of the strength of the spatial correlations and achieves a spatial resolution beyond the limitations of these correlations. 

\end{abstract*}

\section{Introduction}

Numerous complex chemical compounds, including functional groups that are essential for understanding biological processes and organic chemistry, exhibit distinct fingerprints in the mid-infrared (MIR) spectral range \cite{MIR_motiv_2}. These complex molecules absorb low-energy photons in the MIR region, which excites their vibrational modes \cite{MIR_motiv_3, MIR_motiv_4} whereas the excess energy is subsequently released through molecular motion. This characteristic facilitates the implementation of label-free imaging techniques \cite{MIR_motiv_2, MIR_motiv_3}, presenting a significant advantage over fluorophore-based methods that require extensive sample preparation. The rich chemical information provided by this spectral region is leveraged across various fields, including biomedical applications \cite{Offerhaus2019}, spectroscopy\cite{MIR_labelfree_1}, and microscopy \cite{MIR_motiv_5}. 

Despite these advantages, the application of MIR technologies has not been extensively developed. The primary factor contributing to this slow progress is the significant challenge faced by the detection technology. The current stage of MIR detectors lacks high sensitivity, high quantum efficiency, and an optimal signal-to-noise ratio (SNR) which are essential for applications at the single-photon level. Although a limited number of research groups are pursuing advances in this area \cite{Chang:22}, commercially available detectors with single-photon sensitivity in the MIR spectral range are currently inaccessible. Furthermore, MIR detectors typically require a cooling system, which often increases both the size and cost of the equipment.

Quantum imaging with undetected light (QIUL) represents a novel and powerful imaging technique. Through the use of spatially and spectrally correlated photon-pairs, it enables the decoupling of sensing and detection wavelengths, thereby facilitating label-free single-photon imaging in challenging spectral regions, such as the MIR while the detection is carried out at VIS wavelengths \cite{doi:10.1126/sciadv.abd0264, Paterova2020, Krivitsky(2016), Mirko2022, Mirko(2024)}.

Typically, photon-pairs generated by spontaneous parametric down conversion (SPDC) share strong correlations in momentum and position \cite{Walborn2010}. These correlations are present since the moment of creation, after fulfilling energy and momentum conservation \cite{Walborn2010, Lahiri2015}. Using these spatial correlations enabled the implementation of wide-field QIUL, where two different imaging modalities were demonstrated. 
Momentum correlations are exploited if the object is illuminated with the Fourier transform of the crystal plane (2-f imaging configuration) \cite{Sebastian2022, Fuenzalida2023} and position correlations are used when the object is illuminated with the image of the crystal plane (4-f imaging configuration) \cite{MIR_microscopy_Kviatkovsky, marta_position_correlation_resolution}.
In all wide-field implementations of QIUL, the imaging performance is limited by the strength of momentum or position correlations, which degrades the performance of the optical system relative to the diffraction limit of the used optics \cite{marta_position_correlation_resolution, Fuenzalida2022resolutionofquantum, andres_fundamental_res_limit, Viswanathan2021}. Fundamentally, the limitation stems from the limited number of available spatial modes generated in typically used crystals for SPDC, which limits the number of independent image points in the field of view of a wide-field imaging system.

Here, we overcome this boundary by implementing a raster scanning QIUL system in which the sample is illuminated with a focused beam which is scanned across the sample. We demonstrate, that the spatial resolution of our scanning QIUL technique is independent of the strength of the spatial correlations and that our setup achieves a spatial resolution limited only by the employed optical components. 
We show, that the system is well-suited for sensing morphological lipid concentrations and complex functional groups with characteristic fingerprints in the MIR spectral region\cite{Cholesterol_1, Niemann_1}.

\section{Methods and results}
\subsection{Experimental setup}

In this work, we integrate QIUL with scanning microscopy to enhance imaging capabilities and explore novel contrast mechanisms. We utilize SPDC to produce highly non-degenerate signal-idler photon pairs in the VIS and MIR regions, respectively. 

\begin{figure}
\centering\includegraphics[width=\textwidth]{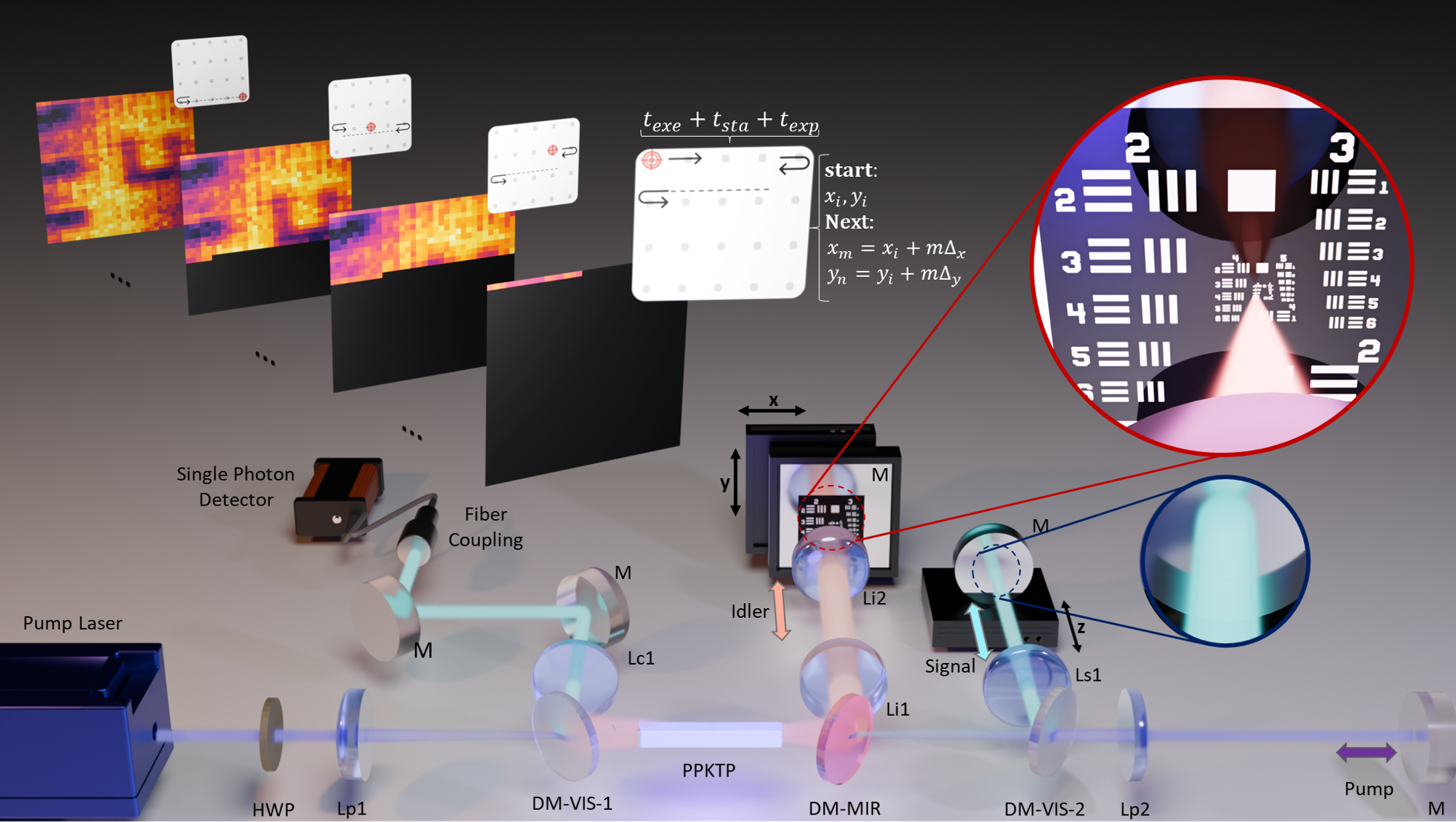}
\caption{Experimental setup for MIR scanning microscopy via QIUL. Double-pass folded nonlinear Michelson interferometer generates signal and idler photons, depicted in cyan and red, respectively, via SPDC. The idler photons interact with the object mounted on an XY translation stage, and the back-reflected photons induce coherence in their signal twin photons, enabling them to interfere at the detection plane. The image reconstruction process shown at the top left is conducted exclusively by the single-pixel detector.}
\label{fig:1}
\end{figure}

A sketch of the experimental setup utilized to perform MIR raster scanning imaging with undetected light is illustrated in Fig. (\ref{fig:1}). 
A 405 nm continuous-wave (CW) laser pumps a 30 mm long ppKTP type-0 nonlinear crystal to generate signal and idler photons via SPDC. The pump beam is focused at the center of the crystal by lens $L_\mathrm{p1}$, with a waist of $w_{\mathrm{p}}=75$ $\mu$m. Subsequently, the pump beam is back-reflected and focused again by lens $L_\mathrm{p2}$; both lenses have the same focal length of 200 mm. Signal (487 nm) and idler (2.4 $\mu$m) photons are generated in the forward and backward paths of the pump beam, and integrated into a non-linear interferometer based on induced coherence \cite{induced_coherence, Ou2020, Mandel1991_2}. All three beams are aligned and recombined at the crystal plane, where the spatial overlap and alignment of the forward- and backward-generated MIR beams induce coherence in their VIS counterpart, enabling interference at the detector plane. This phenomenon allows us to reconstruct an image using the VIS beam, although it has never directly interacted with the sample \cite{Lemos_2014, Sebastian2022, Fuenzalida2023, Leon-Torres(2024)}.

The idler beam is reflected by a dichroic mirror (DM), DM-MIR and focused at the object plane by lenses $L_{\mathrm{i1}}$ and $L_{\mathrm{i2}}$, placed in a 4-f configuration with 100 mm and 50 mm focal lengths, respectively. 
The idler beam probes the object mounted on an Attocube XY motorized translation stage. 
A 0.7 mm thick silicon window is positioned in the idler arm before the object to block any remaining signal photons, ensuring the image reconstruction occurs solely through interaction of the idler photons with the sample.

Pump and signal beams are split by a second dichroic mirror (DM-VIS-2), then collimated and back-reflected to the crystal by lenses $L_{\mathrm{p2}}$ (200 mm focal length) and $L_{\mathrm{s1}}$ (150 mm focal length), respectively. A mirror mounted on a Z-axis motorized translation stage at the end of the signal beam allows precise control of the overall interferometric phase $\Delta \phi$, shown in Eq.(\ref{eq:I}). 

After illuminating the object, the idler photons are back-reflected to the crystal plane to induce coherence in their signal twin photons \cite{Ou2020, Mandel1991_2, induced_coherence}. The probability amplitudes for the generation of forward and backward signal photons interfere when spatial and temporal modes are made indistinguishable, preventing an observer at the detection plane from determining the which-path information of the detected photon \cite{Lemos_2014, Lahiri2015}. The signal photons are redirected for detection through DM-VIS-1 to an Excelitas single-photon detector after being fiber-coupled into a multimode fiber. An ultra-narrow band-pass filter centered at 488 nm with a 1.9 nm bandwidth filters the signal spectrum, while two long-pass filters at 447 nm block the pump beam at the detection plane. A camera is used solely to optimize the alignment of the interferometer by monitoring the contrast of the interference fringes at the crystal plane, while the image reconstruction occurs exclusively through the single-pixel detector. Equation (\ref{eq:I}) shows the photon-count rate at the detection plane \cite{Chekova2016, Lindner:20}.

\begin{equation}
    I \sim 1+|\mathcal{T}(x, y)| \cos{(\Delta \phi + \theta(x, y))},
\label{eq:I}
\end{equation}

where $\Delta \phi$ is the overall interferometric phase that accounts for the pump, signal and idler relative phase difference, $\theta(x, y)$ is the phase acquired by the idler beam when traversing the sample, and $|\mathcal{T}(x, y)|$ is the absolute value of the sample transmission function. 

Measuring the intensity described by Eq. (\ref{eq:I}) at one sample position allows to measure the transmission at this point. To obtain a two-dimensional image, we now scan the object through the fixed idler beam, as shown in the top left of Fig. (\ref{fig:1}). 
The object, placed in the focus of the idler beam, starts from an initial position and is moved step-wise as indicted. After every step, the single-pixel detector records the corresponding intensity counts according to Eq.~\ref{eq:I}. 
For more details on the scanning process and the different time lapses involved, refer to the supplementary material.

In contrast to previous works, our nonlinear interferometer does not rely on spatial correlations between photon pairs. 
In the idler path, the object is illuminated with the image of the crystal plane, a method typically used in position correlation schemes. Meanwhile, in the signal and pump beam paths, the lenses are positioned one focal length from the center of the crystal, as is common in momentum correlation schemes. The crystal length and pump waist were chosen to ensure weak spatial correlations. The key to eliminating dependency on spatial correlations is the point-wise detection scheme, where a single-pixel detector records an intensity value for each raster-scanned point of the object, effectively enforcing spatial correlations through technical means. By eliminating the effect of spatial correlations, the performance of the optical imaging system is constrained only by its classical limitations, namely the diffraction limit \cite{Murphy2013, Webb(1996), Paddock(2000)}.


\subsection{Role of the spatial correlations on QIUL}

A central question our research addresses experimentally is how to eliminate the influence of spatial correlations on the imaging performance of optical systems based on QIUL. The field of QIUL has seen its performance bounded to the classical diffraction limit and the strength of the spatial correlations shared by the photon-pair \cite{Leon-Torres(2024), Fuenzalida2023, Sebastian2022,Lemos_2014}. In all prior studies, the effect of the correlations plays a detrimental role in the performance of the optical system by keeping the spatial resolution above the classical diffraction limit. Only in the ideal case of perfect spatial correlations shared by the photon-pair the optical system does reach the classical diffraction limit; otherwise, the performance is always inferior \cite{Fuenzalida2022resolutionofquantum, MIR_microscopy_Kviatkovsky, andres_fundamental_res_limit, Viswanathan2021, marta_position_correlation_resolution}.

To decrease the influence of spatial correlations from our results, we selected a long nonlinear crystal (30 mm ppKTP with poling period of $\Lambda=5.36$ $\mu$m) and a narrow pump waist of $w_{\mathrm{p}}= 75$ $\mu$m. In the following, we study their impact on the strength of spatial correlations by simulating the joint spatial amplitude (JSA) of the photon-pair.

The JSA is the product between the pump angular distribution $\left| \phi(x_\mathrm{s}, x_\mathrm{i})\right|^2$ and the phase matching function $\left| \psi(x_\mathrm{s}, x_\mathrm{i})\right|^2$, shown in Figs. (\ref{fig:JSA}a - c) for position correlations (see supplementary material) \cite{Viswanathan2021, Chekova2016, Walborn2010}. The geometrical shape of the JSA provides information regarding the strength of the photon-pair correlations. With a diagonal orientation the more elliptical its shape, the sharper the correlations \cite{Chekova2016, Walborn2010}. 
In Fig. (\ref{fig:JSA}c) the JSA exhibits a relatively round shape, implying weak spatial correlations, which is highly detrimental for wide-field QIUL but still allows to implement our scanning technique with high spatial resolution.

\begin{figure}
\centering\includegraphics[width=12cm]{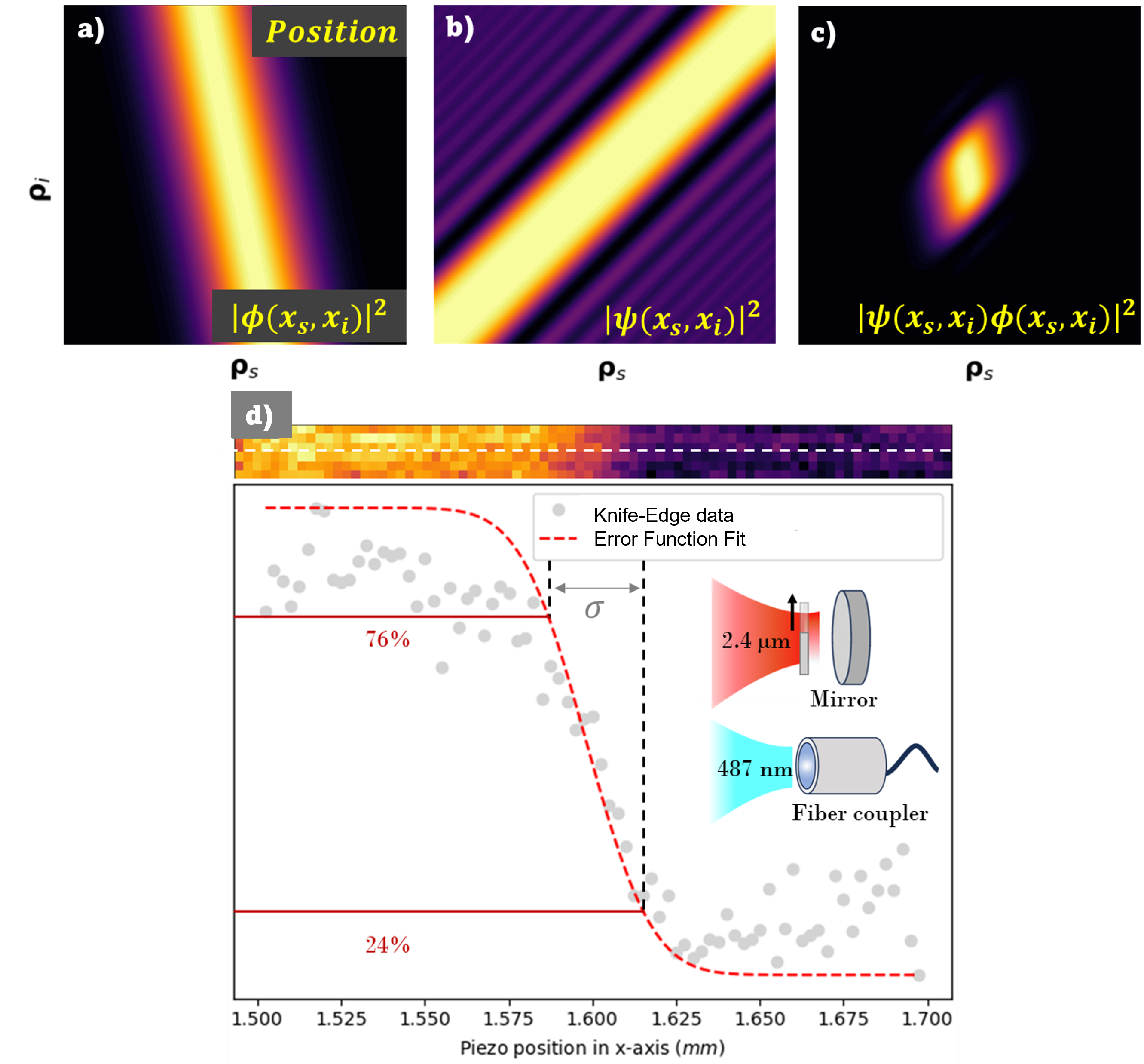}
\caption{Simulation of the joint spatial amplitude (JSA). Panel (a) the pump angular distribution, (b) the phase matching function, and in (c) the JSA in transverse position coordinates are presented, respectively. In panel (d) the Knife-edge measurement of our scanning approach is shown. See supplementary material.}
\label{fig:JSA}
\end{figure}

To demonstrate this, a knife-edge measurement was implemented to estimate the edge spread function (ESF) of our optical system. The top of Fig. (\ref{fig:JSA}d) shows the image of the knife-edge obtained by translating a razor edge along the horizontal axis of the focal plane for the idler beam, we refer to this image with unprocessed data as the raw image. The image is reconstructed by measuring the signal photons at 487 nm. The plotted data was taken from the white dashed line. An error function was fit into the data to estimate the width for which the ESF goes from approximately 24 $\%$ to 76 $\%$ of the maximum value, corresponding to a measurement of the beam radius at 1/e. The knife-edge measurement yields a $\sigma_{\mathrm{scan}}=16.70\ \pm 3.0 \ \mu \mathrm{m}$, obtained after averaging through all lines of the raw image.

To compare the experimentally determined performance of our scanning QIUL system with wide-field approaches, we simulated knife-edge measurements for imaging schemes using position- or momentum correlations. To this end, a sharp edge object is convolved with the point-spread-functions of the optical system based on position and momentum correlations (see supplementary material). For position correlations, the simulation yielded an ESF of $\sigma_{\mathrm{pos}}=174.0$ $\mu$m \cite{marta_position_correlation_resolution}, while for momentum correlations, it yielded $\sigma_{\mathrm{mom}}=506.0$ $\mu$m \cite{Fuenzalida2022resolutionofquantum}. 
This shows, that our scanning approach 
effectively eliminates the influence of spatial correlations on imaging performance.


Our scheme circumvents the use of spatial correlations by implementing a point-wise illumination and detection technique, resulting in an intrinsically correlated output \cite{Gili2022}. An intensity count is associated with each point of the object that is raster scanned, thereby not relying on either position or momentum correlations for image reconstruction. Instead, it relies solely on the phenomenon of induced coherence without induced emission \cite{induced_coherence, Mandel1991_2}, benefiting from distinctly different illumination and detection spectral ranges. 

\subsection{Image formation}

\subsubsection{Raw, amplitude and visibility images comparison}
\begin{figure}
\centering\includegraphics[width=\textwidth]{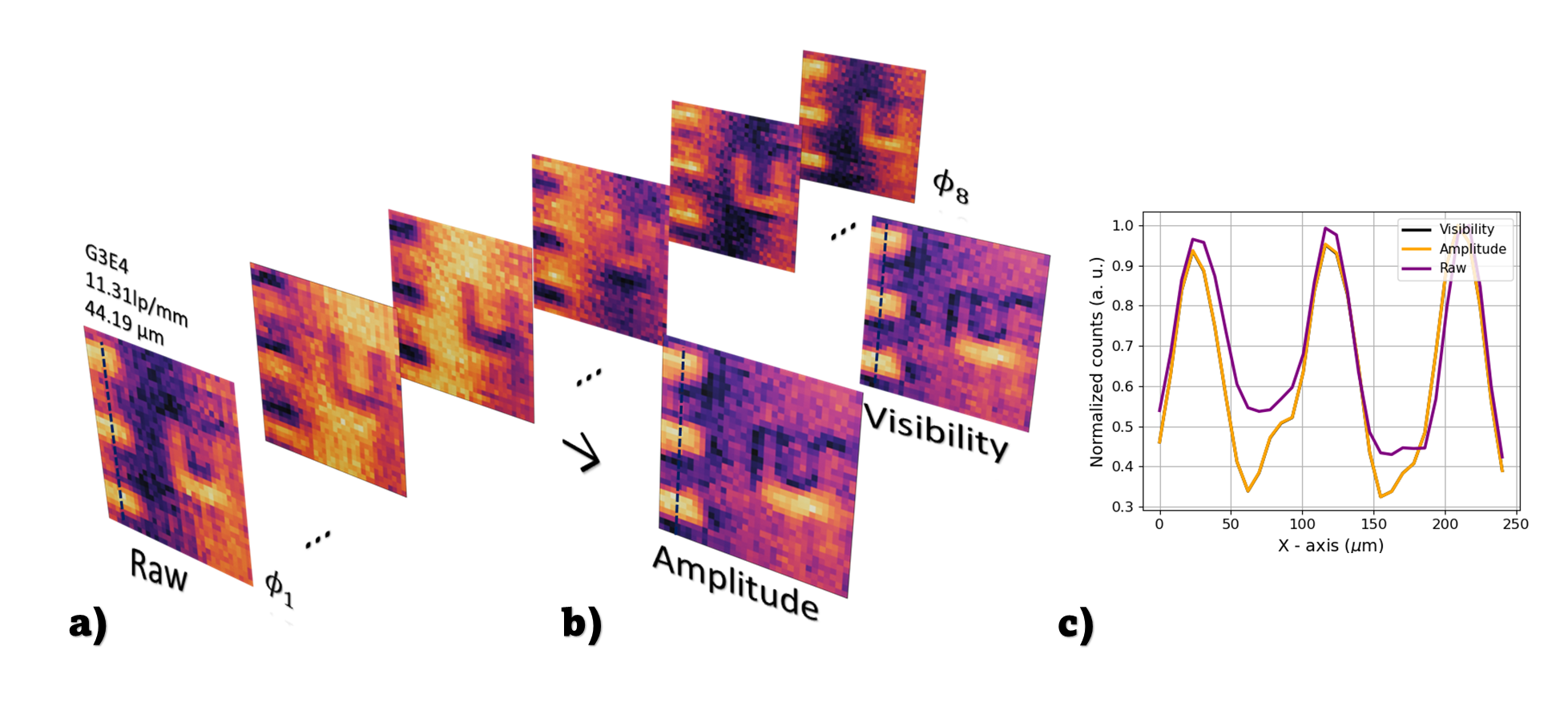}
\caption{Comparison between the raw, amplitude and visibility images. (a) Shows multiple images of element G3E4 acquired by controlling the interferometric phase to calculate the amplitude and visibility images shown in (b). In (c) the transverse plot profile shows a minor improvement in front of the raw image, albeit at the cost of multiple frames and an extended acquisition time.}
\label{fig:raw}
\end{figure}

In QIUL, the measured intensity distribution depends on the phase of the interferometer and thus does not necessarily represent the complex transmission function $\mathcal{T}(x, y)=|\mathcal{T}(x, y)| \exp{[i\theta(x,y)]}$ of the object. On the other hand, $|\mathcal{T}(x, y)|$ is directly proportional to the so-called amplitude $A(x,y)$ and visibility $V(x,y)$ images \cite{marta_position_correlation_resolution, Sebastian2022}, which can be obtained from several raw images measured at different interferometer phases. In the following, we describe how these images are obtained and compare them with the raw images obtained by our scanning approach, shown in Fig. (\ref{fig:raw}). 

Element 4 of group 3 (G3E4) of the USAF resolution target object was imaged, featuring three horizontal bar patterns and the number four, shown in Fig. (\ref{fig:raw} a). Eight raw images were taken at different interferometric phases using the linear translation stage on the signal arm by moving the mirror in small steps along the z-axis ($\lambda_\mathrm{s} /8$). This process enables the reconstruction of the amplitude $A(x,y)=\frac{I_{\mathrm{max}}(x,y)-I_{\mathrm{min}}(x,y)}{2}$ and visibility $V(x,y)=\frac{I_{\mathrm{max}}(x,y)-I_{\mathrm{min}}(x,y)}{I_{\mathrm{max}}(x,y)+I_{\mathrm{min}}(x,y)}$ images seen in Fig. (\ref{fig:raw}b) \cite{Fuenzalida2022resolutionofquantum, Sebastian2022}, where $I_{\mathrm{max}}(x,y)$ and $I_{\mathrm{min}}(x,y)$ are the maximum and minimum intensity values among the eight different images for each scanning position. A plot profile corresponding to the black dashed line in Figs. (\ref{fig:raw}a, b) is shown on Fig. (\ref{fig:raw}c), revealing a minor improvement of contrast in the visibility and amplitude images compared to the raw image.

Unless otherwise stated, all images in this work are raw images obtained at a fixed phase. Given the minor improvement shown in Fig. (\ref{fig:raw}), we concluded that the slight enhancement offered by the visibility or amplitude images does not justify preferring them over the raw image due to the longer image processing time.\\

\subsubsection{Spatial correlations and scanning based imaging schemes}

From a mathematical point of view, the image formation process can be seen as a convolution between the true object and the PSF of the optical system, given as: 
\begin{equation}
\mathrm{Image}(\mathrm{x}, \ \mathrm{y})= \mathrm{Object}(\mathrm{x}, \ \mathrm{y}) \otimes \mathrm{PSF}(\mathrm{x}, \ \mathrm{y}),
\end{equation}

where $\otimes$ denotes a 2D convolution. In order to compare the imaging capabilities of this work with setups that exclusively rely on either position or momentum correlations, the PSFs were simulated for the three cases; see supplementary material. A true object consisting of three horizontal bar patterns and the number four is simulated for different line-width dimensions (45 $\mu$m, 225 $\mu$m, 500 $\mu$m) for the scanning approach, position and momentum correlations, respectively. 

\begin{figure}
\centering\includegraphics[width=\textwidth]{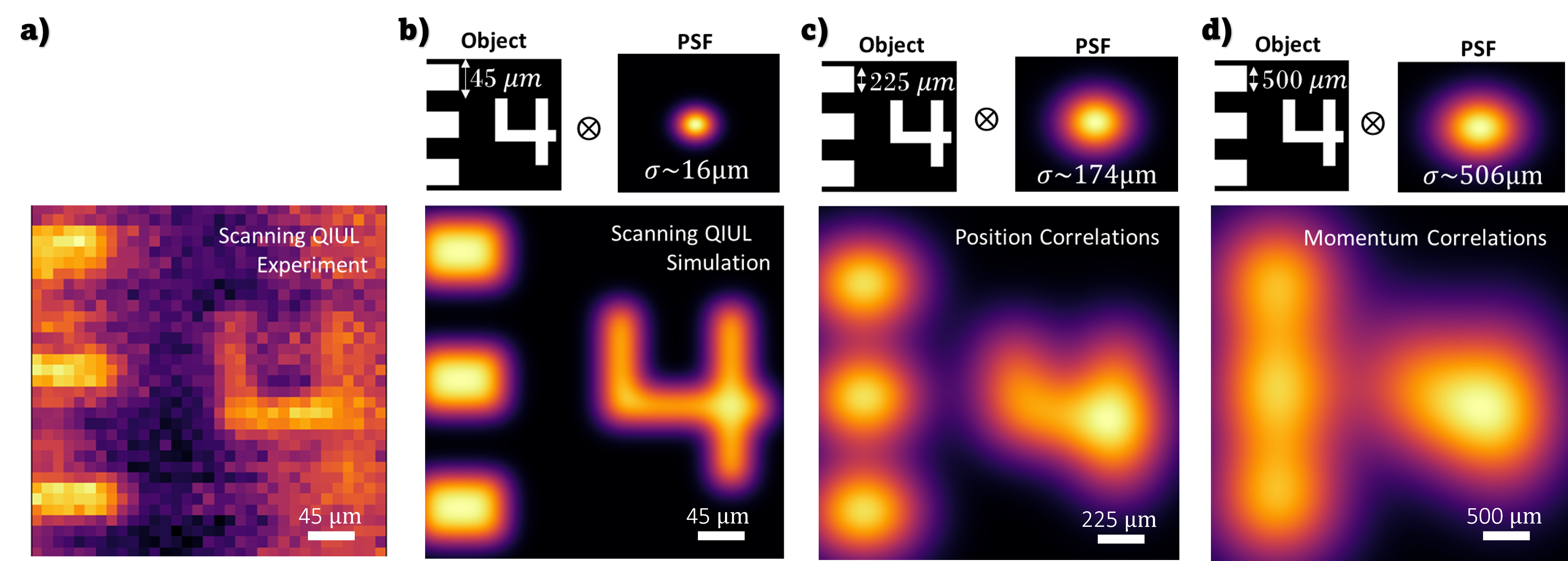}
\caption{Comparison between imaging schemes based on spatial correlations and our scanning approach. (a) element G3E4 of the USAF test target is imaged using our scanning approach, the horizontal bars and number four are clearly resolved. 
A simulated binary object featuring three horizontal bars and the number four was used as a true object to be convolved with the PSF for each case, (b) scanning approach, (c) position correlations, and (d) momentum correlations.}
\label{fig:comparison}
\end{figure}

Figure (\ref{fig:comparison}a) shows the raw image obtained for the G3E4 element with a line-width of 44.19 $\mu$m, the object features are easily resolved. 
Based on the knife-edge measurement results shown in Fig. (\ref{fig:JSA}d), a PSF of 16 $\mu$m radius is simulated for the scanning approach. The result of the 2D-convolution between the true object of 45 $\mu$m features and the PSF for our scanning approach is shown in Fig. (\ref{fig:comparison}b), being in great agreement with our experimental data.

Figure (\ref{fig:comparison}c) features a true object with line-width of 225 $\mu$m, and a PSF of 174 $\mu$m radius for position correlations. In this case, the object features are hardly resolved. Similarly, Fig. (\ref{fig:comparison}d) represents a true object with a line-width of 500 $\mu$m, convolved with a simulated PSF of 506 $\mu$m radius for momentum correlations. As expected, due to the poor momentum correlations shared by the photon-pair, the object features are not resolved. 

If not mentioned otherwise, the PSF radius are calculated at 1/e. Therefore, our scanning QIUL technique eliminates the detrimental effects of spatial correlations and enables diffraction-limited imaging.

\subsection{Characterization of the spatial resolution}

The spatial resolution of the optical system was characterized by different parameters of our imaging system. To this end, its modulation transfer function (MTF), $\mathrm{MTF}=\frac{\pi}{4}*\mathrm{CTF}$, was determined, where $\mathrm{CTF}$ denotes the contrast transfer function given by $\mathrm{CTF}=\frac{I_{\mathrm{max}}-I_{\mathrm{min}}}{I_{\mathrm{max}}+I_{\mathrm{min}}}$ \cite{marta_position_correlation_resolution,Fuenzalida2022resolutionofquantum, Moreau}. The CTF values are obtained from 5 evenly spaced cuts across each image. The $I_{\mathrm{max}}$ and $I_{\mathrm{min}}$ values are extracted from these profiles and used to compute the CTF, which in turn determine the MTF. The average MTF values and standard deviations are then calculated based on the five measurements per image.

\begin{figure}
\centering\includegraphics[width=12.3 cm]{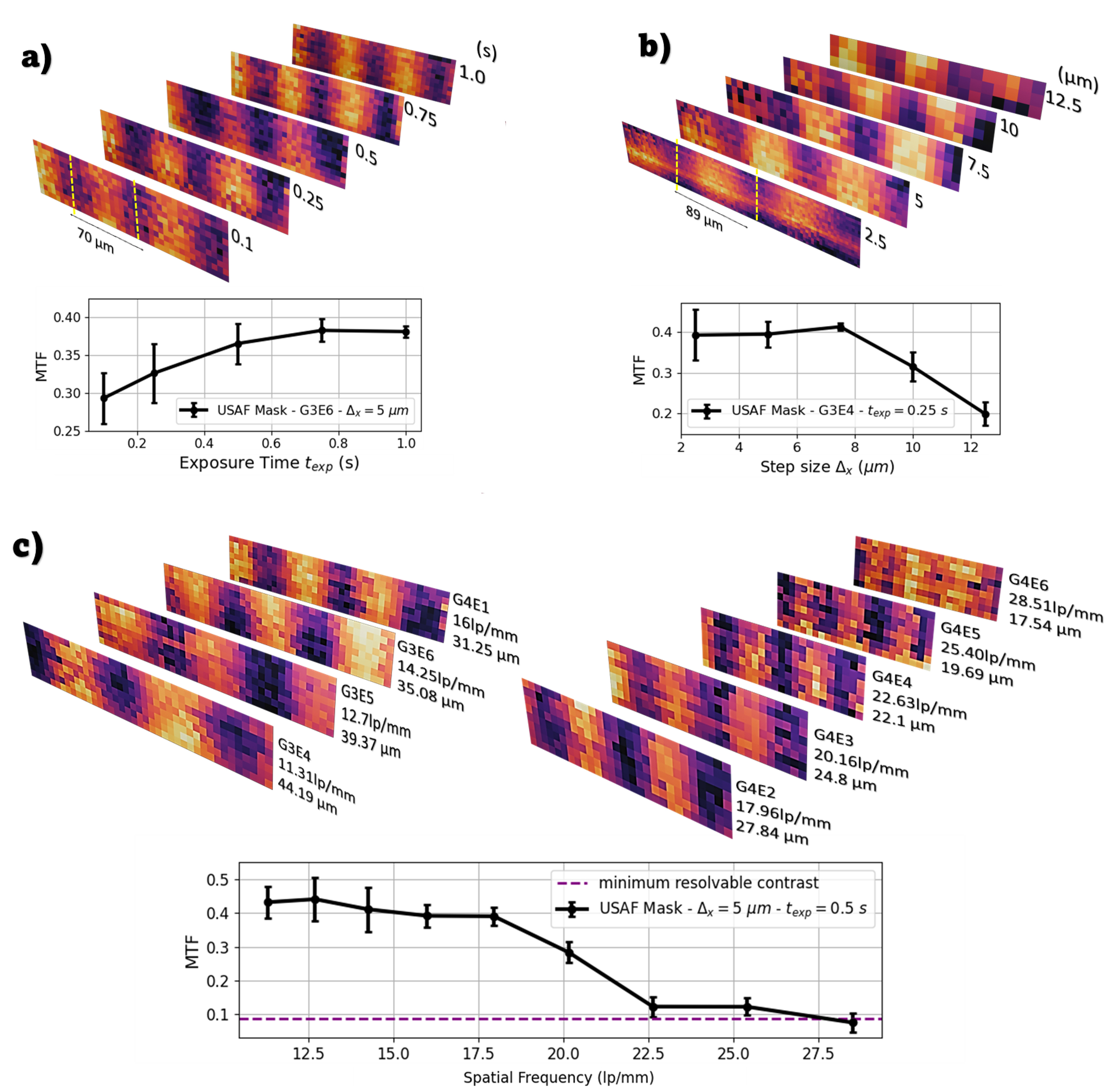}
\caption{Spatial resolution characterization of the optical system through the MTF. (a) MTF as a function of the exposure time and (b) step size. The vertical bars of elements G3E4 to G4E6 of the USAF resolution target were imaged as shown in (c). The highest resolvable spatial frequency corresponds to 25.40 lp/mm.}
\label{fig:MTF}
\end{figure}

First, element 6 of group 3 (G3E6) of the USAF resolution target was imaged at different exposure times, $t_{\mathrm{exp}}= \{0.1,\hspace{0.1cm}  0.25,\hspace{0.1cm}  0.5,\hspace{0.1cm}  0.75,\hspace{0.1cm}  1.0\}$ (s), while maintaining a fixed scanning step size ($\Delta_{\mathrm{x}}=5$ $\mu$m), Fig. (\ref{fig:MTF}a). As shown in the bottom of Fig. (\ref{fig:MTF}a), the MTF values increase with longer exposure times, due to the better signal-to-noise ratio (SNR).

Similarly, while keeping the exposure time constant at 0.25 s, the element G3E4 is imaged at different step sizes $\Delta_{\mathrm{x}}=\{2.5, \hspace{0.1cm} 5, \hspace{0.1cm}  7.5, \hspace{0.1cm}  10, \hspace{0.1cm} 12.5\}$ ($\mu$m), Fig. (\ref{fig:MTF}b). 

As expected, the increase in the step size deteriorates the contrast of the image, due to the inability of lower sampling rates to extract the details of the object. The MTF has been evaluated for each of the images and plotted against the step size, as presented in Fig. (\ref{fig:MTF}b).  

Finally, the highest spatial frequency that can be imaged was determined by imaging the elements G3E4 to G4E6 of the USAF resolution target, shown in Fig. (\ref{fig:MTF}c). A progressive drop in image contrast is seen for higher spatial frequencies due to the inability of the optical system to capture them. All images were taken with a step size of 5 $\mu$m, and exposure time of 0.5 s. The MTF shows that the optical system is capable of resolving up to element G4E5 with 25.40 lp/mm and line-width of 19.7 $\mu$m, bottom of Fig. (\ref{fig:MTF}c).

To determine which elements are resolvable, the Rayleigh criterion is imposed with a 20$\%$ gap between the maximum and minimum for the transverse profile of the vertical elements of the USAF. In Fig. (\ref{fig:MTF}c), we can see that the element G4E6 is not resolvable, its MTF value including the error bar falls below the minimum resolvable contrast given by the Rayleigh criterion.\\


\subsection{Imaging capabilities}

\begin{figure}
\centering\includegraphics[width=12.6cm]{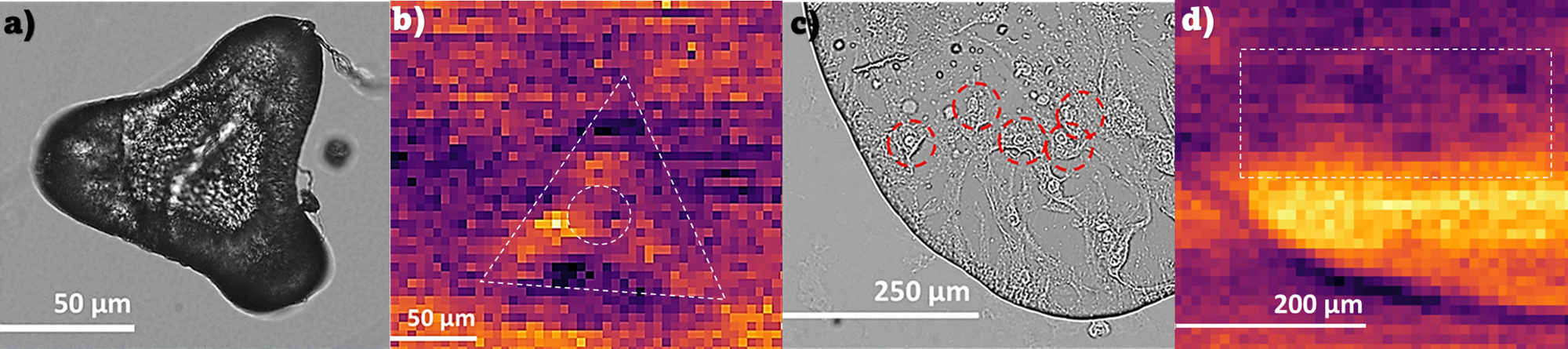}
\caption{(a) and (c) White-light microscope images of Oenothera speciosa pollen samples and a cluster of Fibroblast cells, respectively. (b) and (d) corresponding images obtained by our scanning approach. In (a) the characteristic triangular shape and nucleus structure are recognizable. In (d) the dark spots enclosed in the white-dashed rectangle correspond to the cell nuclei marked by the red-dashed circles on the white-light microscope image.} 
\label{fig:6}
\end{figure}

Our imaging scheme primarily targets the investigation of biological samples, such as complex chemical compounds that vibrate when absorbing MIR photons, resulting in high-contrast imaging. For transparent samples, our approach uses the back-scattered light from the surface of the sample. These images typically exhibit low contrast due to the transparent nature of the specimen and the limited scattering of MIR light.
Figure (\ref{fig:6}a) illustrates white-light microscope images of Oenothera speciosa pollen samples, which are characterized by their triangular shape \cite{pollen}, and a cluster of Fibroblast cells are seen in Fig. (\ref{fig:6}c) \cite{fibroblast}. In Fig. (\ref{fig:6}b) the reconstructed image of the pollen sample is shown, its triangular morphology and nucleolus of the pollen structure is discernible. Additionally, Fig. (\ref{fig:6}d) shows the image of the cluster of fibroblast cells, enclosed by the white-dashed rectangle that corresponds to the regions highlighted by the red-dashed circles.

The samples used in this study were chosen exclusively for illustrative purposes and do not possess any specific scientific or biological relevance.

\section{Discussion}

We have introduced a scanning imaging method based on non-linear interferometry with induced coherence without induced emission. The method raster scans samples with MIR light and carries out detection in the VIS spectral range. Unlike previous works based on QIUL \cite{Mirko(2024), Fuenzalida2023, Leon-Torres(2024), Pearce(2024)}, our approach eliminates the influence of the spatial correlations shared by the photon pair on the imaging performance of the optical system. The only limit to the spatial resolution of the system is the diffraction limit \cite{Murphy2013, Webb(1996)}, therefore our results can be further improved by employing a high-NA objective before the object. The point-wise illumination and detection technique enables a point-by-point correlation between the idler beam at the object plane and the signal beam at the detector plane, associating each scanning position to a gray value, as shown in Fig. (\ref{fig:1}).

To ensure that spatial correlations play no role in our imaging scheme, a relatively long crystal of 30 mm length and narrow pump waist of 75 $\mu$m have been selected. These parameters yield weakly correlated photon pairs in both position and momentum space, as shown in Fig. (\ref{fig:JSA}). Under the same conditions as our experimental implementation, an optical system relying on spatial correlations with undetected light would be unable to resolve two points with a separation distance below 174 $\mu$m and 506 $\mu$m for position and momentum correlations, respectively, see Fig. (\ref{fig:comparison}). In comparison, our approach has a spatial resolution of $16.7 \pm 3.0\ \mu \mathrm{m}$.

The results discussed in Subsections 2.3 and 2.4 demonstrate the capacity of our system to retrieve amplitude information from the USAF resolution target, while Subsection 2.5 highlights its ability to image biological samples like pollen structures and to identify the nuclei of Fibroblast cells. However, the resulting images exhibit limitations primarily due to inaccuracies at positioning the sample at the correct focal plane along the idler beam path. These issues can be mitigated by incorporating a motorized translation stage for precise axial alignment \cite{Murray(2011)thickspecimens, Murphy2013}.

Various parameters can be adjusted to enhance the overall performance of our quantum imaging scheme, including: the use of coated lenses or reflective optics for the idler arm will lead to a better fringe contrast by reducing the losses in the MIR arm. Although the Si-window prevents signal or pump photons from interacting with the object, it causes losses in the idler photon flux, which can be circumvented by better dichroic mirrors. The use of a shorter crystal that allows for a straight forward alignment can significantly improve the fringe contrast. We expect that the use of off-axis parabolic mirrors instead of uncoated lenses improve the overall performance of our imaging system in the idler arm, by avoiding spherical and achromatic aberrations. 

Our results substantially contribute to the field of quantum imaging with undetected light, paving the way for imaging beyond the constraints of spatial correlations while still benefiting from illumination and detection at very distinct wavelength ranges. This innovation bridges the performance gap between quantum imaging with undetected light and standard commercial devices, and relies solely on a single-pixel detector for image reconstruction, making it a cost-effective approach for real-case scenarios.

\begin{backmatter}
\bmsection{Funding} This work was funded by the Deutsche Forschungsgemeinschaft (DFG, German Research Foundation) under Germany´s Excellence Strategy – EXC 2051 – Project-ID 390713860. In addition, this work was supported by the Horizon WIDERA 2021-ACCESS-03-01 grant 101079355 "BioQantSense", from the European Union’s Horizon 2020 Research and Innovation Action under Grant Agreement No. 101113901 (Qu-Test, HORIZON-CL4-2022-QUANTUM-05-SGA) and by a grant funded by the Federal Ministry of Education and Research (QUANTIFISENS - 03RU1U071M, QUANCER - 13N16441).

\bmsection{Acknowledgments} We would like to show our gratitude to W. Grawe for the technical support, and to M. G. Basset, F. Krajinić, R. Sondenheimer, C. Sevilla and V. Kaipalath for fruitful discussions with J.R.L.T. during the course of this research. Furthermore, we thank T. Pajić, A. Jakovljević, C. Credi, and C. Dallari for their help with the biological samples preparation.

\bmsection{Disclosures} The authors declare no conflict of interest.

\bmsection{Data availability} The data that support the findings of this study are available from the corresponding author, J.R.L.T., upon reasonable request.

\bmsection{Supplemental document} See Supplement 1 for supporting content.

\end{backmatter}

\bibliography{Optica-template}

\begin{thebibliography}{10}
\newcommand{\enquote}[1]{``#1''}

\bibitem{MIR_motiv_2}
M.~Hermes, R.~B. Morrish, L.~Huot, \emph{et~al.}, \enquote{Mid-ir hyperspectral imaging for label-free histopathology and cytology,} {\protect\JournalTitle{Journal of Optics (United Kingdom)}} \textbf{20} (2018).

\bibitem{MIR_motiv_3}
L.~Shi, X.~Liu, L.~Shi, \emph{et~al.}, \enquote{Mid-infrared metabolic imaging with vibrational probes,} {\protect\JournalTitle{Nature Methods}} \textbf{17}, 844--851 (2020).

\bibitem{MIR_motiv_4}
Y.~Zhao, S.~Kusama, Y.~Furutani, \emph{et~al.}, \enquote{High-speed scanless entire bandwidth mid-infrared chemical imaging,} {\protect\JournalTitle{Nature Communications}} \textbf{14} (2023).

\bibitem{Offerhaus2019}
H.~L. Offerhaus, S.~E. Bohndiek, and A.~R. Harvey, \enquote{Hyperspectral imaging in biomedical applications,} {\protect\JournalTitle{Journal of Optics (United Kingdom)}} \textbf{21} (2019).

\bibitem{MIR_labelfree_1}
S.~Clède, C.~Policar, and C.~Sandt, \enquote{Fourier transform infrared (ft-ir) spectromicroscopy to identify cell organelles: Correlation with fluorescence staining in mcf-7 breast cancer cells,} {\protect\JournalTitle{Applied Spectroscopy}} \textbf{68}, 113--117 (2014).

\bibitem{MIR_motiv_5}
A.~Ebner, P.~Gattinger, I.~Zorin, \emph{et~al.}, \enquote{Diffraction-limited hyperspectral mid-infrared single-pixel microscopy,} {\protect\JournalTitle{Scientific Reports}} \textbf{13} (2023).

\bibitem{Chang:22}
J.~Chang, J.~W.~N. Los, R.~Gourgues, \emph{et~al.}, \enquote{Efficient mid-infrared single-photon detection using superconducting nbtin nanowires with high time resolution in a gifford-mcmahon cryocooler,} {\protect\JournalTitle{Photon. Res.}} \textbf{10}, 1063--1070 (2022).

\bibitem{doi:10.1126/sciadv.abd0264}
I.~Kviatkovsky, H.~M. Chrzanowski, E.~G. Avery, \emph{et~al.}, \enquote{Microscopy with undetected photons in the mid-infrared,} {\protect\JournalTitle{Science Advances}} \textbf{6}, eabd0264 (2020).

\bibitem{Paterova2020}
A.~V. Paterova, S.~M. Maniam, H.~Yang, \emph{et~al.}, \enquote{Hyperspectral infrared microscopy with visible light,}  (2020).

\bibitem{Krivitsky(2016)}
D.~A. Kalashnikov, A.~V. Paterova, S.~P. Kulik, and L.~A. Krivitsky, \enquote{Infrared spectroscopy with visible light,} {\protect\JournalTitle{Nature Photonics}} \textbf{10}, 98--101 (2016).

\bibitem{Mirko2022}
M.~Kutas, B.~E. Haase, F.~Riexinger, \emph{et~al.}, \enquote{Quantum sensing with extreme light,} {\protect\JournalTitle{Advanced Quantum Technologies}} \textbf{5} (2022).

\bibitem{Mirko(2024)}
M.~Kutas, F.~Riexinger, J.~Klier, \emph{et~al.}, \enquote{Terahertz quantum imaging,}  (2024).

\bibitem{Walborn2010}
S.~P. Walborn, C.~H. Monken, S.~Pádua, and P.~H.~S. Ribeiro, \enquote{Spatial correlations in parametric down-conversion,} {\protect\JournalTitle{Physics Reports}} \textbf{495}, 87--139 (2010).

\bibitem{Lahiri2015}
M.~Lahiri, R.~Lapkiewicz, G.~B. Lemos, and A.~Zeilinger, \enquote{Theory of quantum imaging with undetected photons,} {\protect\JournalTitle{Physical Review A - Atomic, Molecular, and Optical Physics}} \textbf{92} (2015).

\bibitem{Sebastian2022}
S.~Töpfer, M.~G. Basset, J.~Fuenzalida, \emph{et~al.}, \enquote{Quantum holography with undetected light,} {\protect\JournalTitle{Sci. Adv}} \textbf{8}, 4301 (2022).

\bibitem{Fuenzalida2023}
J.~Fuenzalida, M.~G. Basset, S.~Töpfer, \emph{et~al.}, \enquote{Experimental quantum imaging distillation with undetected light,} {\protect\JournalTitle{Science Advances}} \textbf{9} (2023).

\bibitem{MIR_microscopy_Kviatkovsky}
I.~Kviatkovsky, H.~Chrzanowski, and S.~Ramelow, \enquote{Mid-infrared microscopy via position correlations of undetected photons,} {\protect\JournalTitle{Optics Express}} \textbf{30} (2022).

\bibitem{marta_position_correlation_resolution}
M.~Gilaberte~Basset, R.~Sondenheimer, J.~Fuenzalida, \emph{et~al.}, \enquote{Experimental analysis of image resolution of quantum imaging with undetected light through position correlations,} {\protect\JournalTitle{Phys. Rev. A}} \textbf{108}, 052610 (2023).

\bibitem{Fuenzalida2022resolutionofquantum}
J.~Fuenzalida, A.~Hochrainer, G.~B. Lemos, \emph{et~al.}, \enquote{Resolution of {Q}uantum {I}maging with {U}ndetected {P}hotons,} {\protect\JournalTitle{{Quantum}}} \textbf{6}, 646 (2022).

\bibitem{andres_fundamental_res_limit}
A.~Vega, E.~A. Santos, J.~Fuenzalida, \emph{et~al.}, \enquote{Fundamental resolution limit of quantum imaging with undetected photons,} {\protect\JournalTitle{Phys. Rev. Res.}} \textbf{4}, 033252 (2022).

\bibitem{Viswanathan2021}
B.~Viswanathan, G.~B. Lemos, and M.~Lahiri, \enquote{Resolution limit in quantum imaging with undetected photons using position correlations,} {\protect\JournalTitle{Optics Express}} \textbf{29}, 38185 (2021).

\bibitem{Cholesterol_1}
U.~Gupta, V.~Singh, V.~Kumar, and Y.~Khajuria, \enquote{Spectroscopic studies of cholesterol: Fourier transform infra-red and vibrational frequency analysis,} {\protect\JournalTitle{Materials Focus}} \textbf{3}, 211--217 (2014).

\bibitem{Niemann_1}
J.~E. Vance and B.~Karten, \enquote{Niemann-pick c disease and mobilization of lysosomal cholesterol by cyclodextrin,} {\protect\JournalTitle{Journal of Lipid Research}} \textbf{55}, 1609--1621 (2014).

\bibitem{induced_coherence}
X.~Y. Zou, L.~J. Wang, and L.~Mandel, \enquote{Induced coherence and indistinguishability in optical interference,} {\protect\JournalTitle{Phys. Rev. Lett.}} \textbf{67}, 318--321 (1991).

\bibitem{Ou2020}
Z.~Y. Ou and X.~Li, \enquote{Quantum su(1,1) interferometers: Basic principles and applications,} {\protect\JournalTitle{APL Photonics}} \textbf{5} (2020).

\bibitem{Mandel1991_2}
I.~J. Wang, X.~Y. Zou, and L.~Mandel, \enquote{Induced coherence without induced emission,} {\protect\JournalTitle{Physical Review A}} \textbf{44}, 4615--4622 (1991).

\bibitem{Lemos_2014}
G.~B. Lemos, V.~Borish, G.~D. Cole, \emph{et~al.}, \enquote{Quantum imaging with undetected photons,} {\protect\JournalTitle{Nature}} \textbf{512}, 409--412 (2014).

\bibitem{Leon-Torres(2024)}
J.~R. León-Torres, F.~Krajinic, M.~Kumar, \emph{et~al.}, \enquote{Off-axis holographic imaging with undetected light,} {\protect\JournalTitle{Optics Express}}  (2024).

\bibitem{Chekova2016}
M.~V. Chekhova and Z.~Y. Ou, \enquote{Nonlinear interferometers in quantum optics,} {\protect\JournalTitle{Advances in Optics and Photonics}} \textbf{8}, 104 (2016).

\bibitem{Lindner:20}
C.~Lindner, S.~Wolf, J.~Kiessling, and F.~K\"{u}hnemann, \enquote{Fourier transform infrared spectroscopy with visible light,} {\protect\JournalTitle{Opt. Express}} \textbf{28}, 4426--4432 (2020).

\bibitem{Murphy2013}
D.~B. Murphy and M.~W. Davidson, \emph{Fundamentals of light microscopy and electronic imaging} (Wiley-Blackwell, 2013).

\bibitem{Webb(1996)}
W.~H. Robert, \enquote{Confocal optical microscopy,} {\protect\JournalTitle{Reports on Progress in Physics}} \textbf{59}, 427--471 (1996).

\bibitem{Paddock(2000)}
P.~W. Stephen, \enquote{Principles and practices of laser scanning confocal microscopy,} {\protect\JournalTitle{Molecular Biotechnology}} \textbf{16} (2000).

\bibitem{Gili2022}
V.~F. Gili, C.~Piccinini, M.~Safari~Arabi, \emph{et~al.}, \enquote{Experimental realization of scanning quantum microscopy,} {\protect\JournalTitle{Applied Physics Letters}} \textbf{121}, 104002 (2022).

\bibitem{Moreau}
P.-A. Moreau, E.~Toninelli, P.~A. Morris, \emph{et~al.}, \enquote{Resolution limits of quantum ghost imaging,} {\protect\JournalTitle{Optics Express}} \textbf{26}, 7528 (2018).

\bibitem{pollen}
B.~Zlatkov, S.~Beshkov, and T.~Ganeva, \enquote{Oenothera speciosa versus macroglossum stellatarum: killing beauty,} {\protect\JournalTitle{Arthropod-Plant Interactions}} \textbf{12}, 395–400 (2018).

\bibitem{fibroblast}
K.~Eberhardt, C.~Beleites, S.~Marthandan, \emph{et~al.}, \enquote{Raman and infrared spectroscopy distinguishing replicative senescent from proliferating primary human fibroblast cells by detecting spectral differences mainly due to biomolecular alterations,} {\protect\JournalTitle{Analytical Chemistry}} \textbf{89}, 2937--2947 (2017). PMID: 28192961.

\bibitem{Pearce(2024)}
E.~Pearce, O.~Wolley, S.~P. Mekhail, \emph{et~al.}, \enquote{Single-frame transmission and phase imaging using off-axis holography with undetected photons,} {\protect\JournalTitle{Scientific Reports}} \textbf{14} (2024).

\bibitem{Murray(2011)thickspecimens}
J.~M. Murray, \enquote{Methods for imaging thick specimens: Confocal microscopy, deconvolution, and structured illumination,} {\protect\JournalTitle{Cold Spring Harbor Protocols}} \textbf{6}, 1399--1437 (2011).

\end{thebibliography}


\begin{thebibliography}{1}
\newcommand{\enquote}[1]{``#1''}

\bibitem{Viswanathan:21}
B.~Viswanathan, G.~B. Lemos, and M.~Lahiri, \enquote{Resolution limit in quantum imaging with undetected photons using position correlations,} {\protect\JournalTitle{Opt. Express}} \textbf{29}, 38185--38198 (2021).

\bibitem{Chekova2016}
M.~V. Chekhova and Z.~Y. Ou, \enquote{Nonlinear interferometers in quantum optics,} {\protect\JournalTitle{Advances in Optics and Photonics}} \textbf{8}, 104 (2016).

\bibitem{Walborn2010}
S.~P. Walborn, C.~H. Monken, S.~Pádua, and P.~H.~S. Ribeiro, \enquote{Spatial correlations in parametric down-conversion,} {\protect\JournalTitle{Physics Reports}} \textbf{495}, 87--139 (2010).

\bibitem{Heintzmann(2006)}
R.~Heintzmann, \enquote{Band limit and appropriate sampling in microscopy,}  (2006).

\bibitem{Heintzmann(2007)}
R.~Heintzmann and C.~J. Sheppard, \enquote{The sampling limit in fluorescence microscopy,} {\protect\JournalTitle{Micron}} \textbf{38}, 145--149 (2007).

\bibitem{Murphy2013}
D.~B. Murphy and M.~W. Davidson, \emph{Fundamentals of light microscopy and electronic imaging} (Wiley-Blackwell, 2013).

\bibitem{marta_position_correlation_resolution}
M.~Gilaberte~Basset, R.~Sondenheimer, J.~Fuenzalida, \emph{et~al.}, \enquote{Experimental analysis of image resolution of quantum imaging with undetected light through position correlations,} {\protect\JournalTitle{Phys. Rev. A}} \textbf{108}, 052610 (2023).

\bibitem{Fuenzalida2022resolutionofquantum}
J.~Fuenzalida, A.~Hochrainer, G.~B. Lemos, \emph{et~al.}, \enquote{Resolution of {Q}uantum {I}maging with {U}ndetected {P}hotons,} {\protect\JournalTitle{{Quantum}}} \textbf{6}, 646 (2022).

\end{thebibliography}

\end{document}


\maketitle

\section{Joint Spatial Amplitude (JSA)}

In this section we show the expression for the JSA using the same approach highlighted in \cite{Viswanathan:21}. The JSA is defined as the product of the pump angular distribution $\left| \phi(q_{\mathrm{s}}, q_{\mathrm{i}})\right|^2$ and the phase matching function $\left| \psi(q_{\mathrm{s}}, q_{\mathrm{i}}) \right|^2$ in the transverse momentum coordinates. The pump angular distribution follows a Gaussian profile, described by \cite{ Chekova2016, Walborn2010}:

\begin{equation}
    \left| \phi(q_{\mathrm{s}}, q_{\mathrm{i}})\right|^2 = \exp{\left( - \frac{\left| q_{\mathrm{s}} + q_{\mathrm{i}}\right|^2}{4} w_{\mathrm{p}}^2 \right)}, 
    \label{eq:S1}
\end{equation}

\noindent where $w_{\mathrm{p}}$ is the pump waist inside the crystal and $q_{\mathrm{s}},\ q_{\mathrm{i}}$ represent the transverse momenta of the signal and idler photons, respectively. These momenta describe the spatial frequency components of the twin photons generated in the spontaneous parametric down conversion (SPDC) process. The standard phase matching function for the SPDC can be approximated by the following Gaussian function:

\begin{equation}
   \left| \psi(q_{\mathrm{s}}, q_{\mathrm{i}}) \right|^2 = \exp{\left( - \frac{L\lambda_{\mathrm{p}} \lambda_{\mathrm{s}}}{8\pi\lambda_{\mathrm{i}}} \left| q_{\mathrm{s}} - \frac{\lambda_{\mathrm{i}}}{\lambda_{\mathrm{s}}}q_{\mathrm{i}} \right|^2 \right)},
\end{equation}

\noindent where $L$ is the length of the nonlinear crystal, and $\lambda_{\mathrm{p,s,i}}$ are the pump, signal (detected) and idler (undetected) wavelengths. Therefore, the JSA in the momentum space becomes:

\begin{equation}
    \text{JSA}(q_{\mathrm{s}}, q_{\mathrm{i}}) \propto \exp{\left( - \frac{\left| q_{\mathrm{s}} + q_{\mathrm{i}}\right|^2}{4} w_\mathrm{p}^2 \right)} \exp{\left( - \frac{L\lambda_{\mathrm{p}} \lambda_{\mathrm{s}}}{8\pi\lambda_{\mathrm{i}}} \left| q_{\mathrm{s}} - \frac{\lambda_{\mathrm{i}}}{\lambda_{\mathrm{s}}}q_{\mathrm{i}} \right|^2 \right)}.
\label{eq:S3}
\end{equation}

\begin{figure}[ht!]
\centering\includegraphics[width=\textwidth]{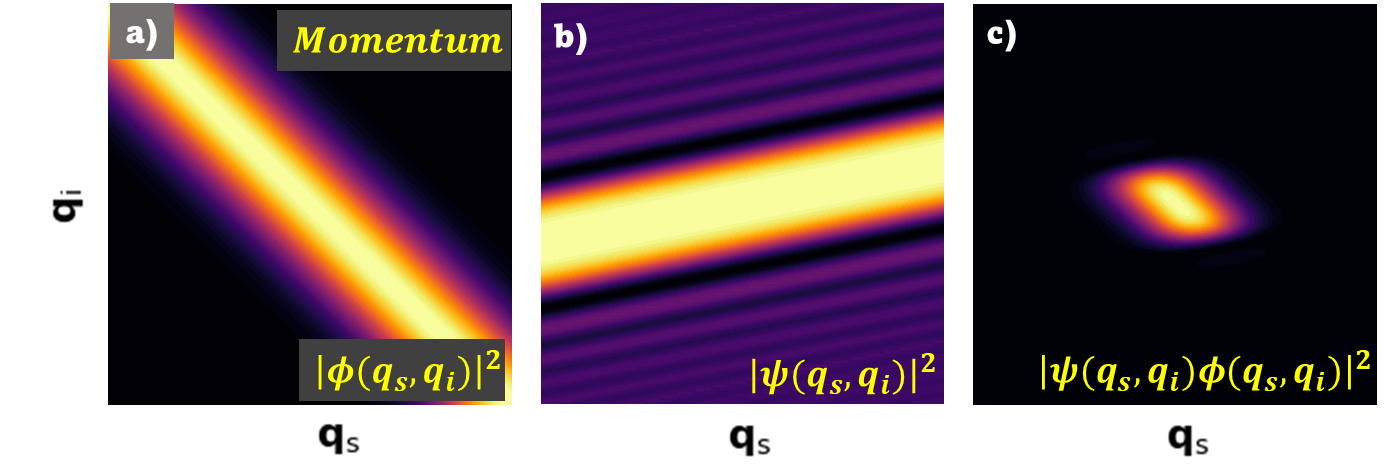}
\caption{Simulation of the joint spatial amplitude (JSA). Panel (a) the pump angular distribution, (b) the phase matching function, and in (c) the JSA in transverse momentum coordinates are presented, respectively. }
\label{fig:S1}
\end{figure}

Equations (\ref{eq:S1}) - (\ref{eq:S3}) are simulated and shown in Fig. (\ref{fig:S1}).

Taking into account the normalization condition of the quantum state generated by the SPDC, the JSA in terms of transverse position coordinates $(\rho_\mathrm{s}, \rho_\mathrm{i})$ becomes: 

\begin{align}
\text{JSA}(\rho_\mathrm{s}, \rho_\mathrm{i}) =
\frac{8}{\pi L w_\mathrm{p}^2 (\lambda_\mathrm{s} + \lambda_\mathrm{i})}
\exp \left[ -\frac{4\pi}{L (\lambda_\mathrm{s} + \lambda_\mathrm{i})} |\rho_\mathrm{s} - \rho_\mathrm{i}|^2 \right] \notag \\
\exp \left[ -\frac{2}{w_\mathrm{p}^2 (\lambda_\mathrm{i} + \lambda_\mathrm{s})^2} |\lambda_\mathrm{i} \rho_\mathrm{s} + \lambda_\mathrm{s} \rho_\mathrm{i}|^2 \right].
\end{align}

where the second term represents the phase matching function, and the pump angular distribution is given by the third term, shown in Figs. (2a) - (2c) of the main text.

\section{Point spread function (PSF)}

The image produced by an optical imaging system is mathematically described as the convolution of the object with the PSF of the optical system shown in Eq. (S5), which characterizes its spatial resolution and response to a point source \cite{Heintzmann(2006), Heintzmann(2007), Murphy2013}.

\begin{equation}
\mathrm{Image}(\mathrm{x}, \ \mathrm{y})= \mathrm{Object}(\mathrm{x}, \ \mathrm{y}) \otimes \mathrm{PSF}(\mathrm{x}, \ \mathrm{y}).
\end{equation}

For the case of an imaging system that relies on quantum imaging with undetected light by position correlation, the PSF is given by \cite{marta_position_correlation_resolution}:

\begin{equation}
\text{PSF}_{\text{pos.corr.}}(\rho_c) =
\exp \left\{ - 2 \
\frac{\left[ 2\pi w_\mathrm{p}^2 (\lambda_\mathrm{s} + \lambda_\mathrm{i}) - \lambda_\mathrm{s} \lambda_\mathrm{i} L \right]^2 }
{2\pi w_\mathrm{p}^4 (\lambda_\mathrm{s} + \lambda_\mathrm{i})^3 L + w_\mathrm{p}^2 \lambda_\mathrm{s}^2 (\lambda_\mathrm{s} + \lambda_\mathrm{i})^2 L^2}
\frac{\rho_\mathrm{c}^2}{M_\mathrm{s}^2} \right\},
\end{equation}

\noindent where $\rho_\mathrm{c}$ denotes the transverse position of the detected photon on the camera plane, given by $\rho_\mathrm{c} = \sqrt{x^2+y^2}$. $M_\mathrm{s}$ is the total magnification in the signal beam path. It accounts for how the detected spatial information of the photon is mapped onto the camera. For our parameters $M_{\mathrm{s}}=\frac{f_{L_{\mathrm{c2}}}}{f_{L_{\mathrm{c1}}}}$, where the focal lengths take the following values $f_{L_{\mathrm{c1}}}=150 \ \mathrm{mm}$  and $f_{L_{\mathrm{c2}}}=200 \ \mathrm{mm}$. \\

We assume that the PSF for an optical system based on momentum correlations follows a Gaussian profile with width measured at 1/e given by \cite{Fuenzalida2022resolutionofquantum}:

\begin{equation}
\sigma_{\text{mom.corr.}} = \frac{f_{L_{\mathrm{i2}}} \lambda_\mathrm{i}}{\sqrt{2} \pi w_\mathrm{p}},
\end{equation}

\begin{equation}
\text{PSF}_{\text{mom.corr.}}(\rho_\mathrm{c}) =
\exp \left ( - 2 \
\frac{\rho^{2}_{\mathrm{c}}}{\sigma_{\text{mom.corr.}}} \right ),
\end{equation}

\noindent where $f_{L_{i2}}$ is the focal length of the lens in front of the object plane. $\sigma_{\text{mom.corr.}}$ is the distance on the camera plane where visibility increases from approximately 24\% to 76\% of its maximum value. Experimentally, it is determined by the knife-edge response \cite{Fuenzalida2022resolutionofquantum}.\\

In contrast to the wide-field quantum imaging with undetected light, in the raster scanning case the PSF is not limited by the spatial correlations. It is given by the classical diffraction theory, being characterized by the first zero of the Airy disk \cite{Murphy2013}:

\begin{equation}
    r = 0.61\frac{ \lambda_\mathrm{i}}{\text{NA}},
\end{equation}

\begin{equation}
\text{PSF}_{\text{scan.}}(\rho_\mathrm{c}) =
\exp \left ( - \frac{2}{\ln{2}}
\frac{\rho^{2}_{\mathrm{c}}}{r} \right ),
\end{equation}

\noindent where NA is the numerical aperture of the focusing lens ($L_{\mathrm{i2}}$). In our scanning approach the beam radius (at 1/e) obtained through the knife-edge measurement is equivalent to the quantity $\sigma = r  \ln{2} $. 

\begin{figure}[ht!]
\centering\includegraphics[width=\textwidth]{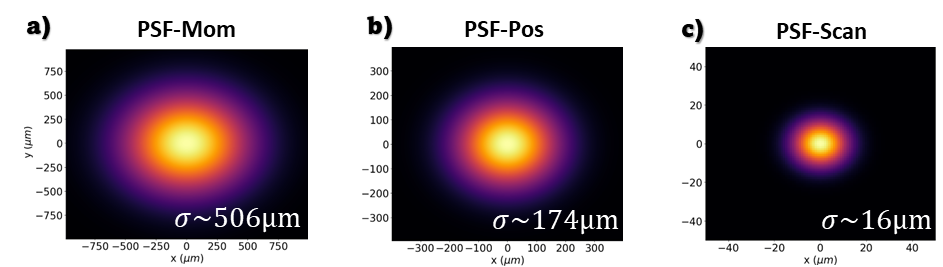}
\caption{The simulated PSFs for an optical imaging system relying on momentum (left), position (center) correlations and our scanning approach (right). }
\label{fig:S2}
\end{figure}

Figure (\ref{fig:S2}) showcases the PSFs for each case, the sigma value corresponds to the radius of the PSF (at 1/e). For an optical system relying on momentum correlations (a), position correlations (b) and raster scanning imaging (c).

\section{Edge spread function (ESF)}

In Sec. (2.2) of the main text, we calculated the ESF for both cases of position and momentum correlations. In the following, we clarify the steps taken for this simulation. 

An edge object is simulated by:

\begin{equation}
    \mathrm{edge}(x, y) =
    \begin{cases} 
    1 & \text{if } -\infty \leq x < 0 \ \text{ and } \ y \geq 0 \\
    0 & \text{if } 0 \leq x \leq +\infty \ \text{ and} \ y \geq 0.  
    \end{cases}
\end{equation}

\begin{figure}[ht!]
\centering\includegraphics[width=\textwidth]{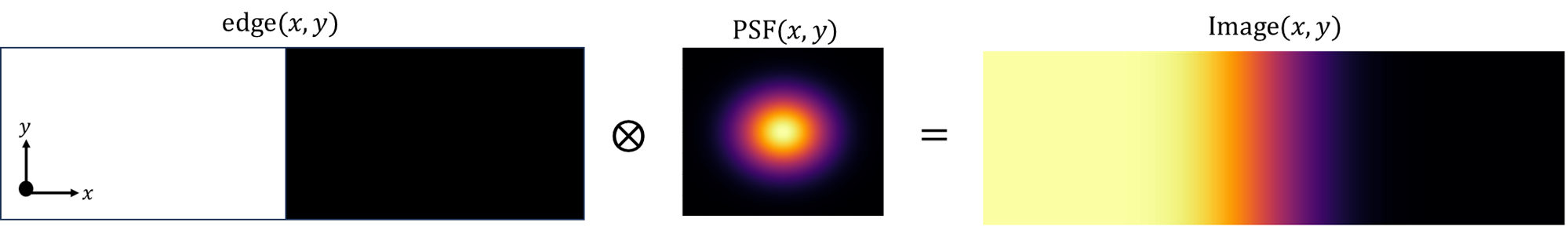}
\caption{2D convolution process between the edge true object and the simulated PSF of the optical system.}
\label{fig:s3}
\end{figure}

The edge object is convolved with the PSF for every case obtaining the simulated reconstructed image from which the ESF is extracted. Fig. (\ref{fig:s3}) illustrates the process.

\section{Image processing}

In this Section, we offer more detail in the technical aspects of our image reconstruction process.
The top left of Fig. (1), in the main text, illustrates graphically the raster scanning process of our setup. 
The object, placed in the focal plane of the idler beam, starts from an initial position ($x_{\mathrm{i}}$, $y_{\mathrm{i}}$). The motorized translation stage (MTS) moves along the x-axis in a constant step size ($\Delta_{\mathrm{x}}$), updating the position to $x_{m}=x_{\mathrm{i}}+m \Delta_{\mathrm{x}}$ for $ m \in \{0, 1, \ldots, M\}$, where M is the total number of x-axis steps. Upon reaching the x-axis end, it returns to the initial position and moves one step down the y-axis to $y_{n}=y_{\mathrm{i}}+n \Delta_{\mathrm{y}}$, where $\Delta_{\mathrm{y}}$ is the y-axis step size, and $n \in \{0, 1, \ldots, N\}$, with N being the total number of y-axis steps. 

These values are stored in a matrix and converted to an 8-bit image with gray values from 0 to 255. An optimal sampling rate ($\Delta_{\mathrm{x}}$, $\Delta_{\mathrm{y}}$) ensures a correct setup operation \cite{Heintzmann(2006), Heintzmann(2007)}.

Each step involves three processes. First, the execution process ($t_{\mathrm{exe}}$) measures the time from the moment the MTS is assigned a new position until the moment the position is reached.

This time depends on the step size used for scanning; a larger $\Delta_{\mathrm{x}}$ results in a longer $t_{\mathrm{exe}}$. For our measurements, $t_{\mathrm{exe}}$ is approximately 100 ms.

Second, the stabilization process introduces a time window ($t_{\mathrm{sta}}$) from the instant the MTS reaches its new position until the measurement is taken. A value of $t_{\mathrm{sta}}= 0.5$ s was used for all measurements. 

And third, the intensity measurement process. It records the counts in the single pixel detector using a time tagger; this time lapse is the exposure time ($t_{\mathrm{exp}}$).

The images in Fig. (3) of the main text were collected using the following parameters: 
$\Delta_{\mathrm{x}}=\Delta_{\mathrm{y}}=7.5$ ${\mu}$m, $t_{\mathrm{exp}}=0.25$ s, total number of steps ($M\times N$=1089), and total execution time $M\times N\times (t_{\mathrm{exe}}+t_{\mathrm{sta}}+t_{\mathrm{exp}})= 15.4$ min.

\bibliography{sa-supplemental-document-template}
